\documentclass[aps,pre,reprint,nofootinbib]{revtex4-2}
\usepackage{amsmath,amssymb,bm,amsthm,booktabs,graphicx,hyperref}
\hypersetup{colorlinks=true,linkcolor=blue,citecolor=blue,urlcolor=blue}
\newtheorem{theorem}{Theorem}
\newtheorem{proposition}{Proposition}
\newtheorem{corollary}{Corollary}
\newcommand{\T}{\mathbb T}
\newcommand{\R}{\mathbb R}
\newcommand{\eps}{\varepsilon}
\newcommand{\meas}{\operatorname{meas}}
\newcommand{\OO}{\mathcal O}
\newcommand{\norm}[1]{\left\lVert#1\right\rVert}
\begin{document}
\title{Regular motion in strongly perturbed dynamical systems}
\author{Marco Frasca}
\affiliation{Independent Researcher, Rome, Italy}
\date{\today}
\begin{abstract}
Strong perturbations on dynamical systems are generally deemed as untractable unless we use numerical methods. It is a rather general expectation that chaotic dynamics could set in impeding any possible analytical treatment. We show a corollary, derived from the Kolmogorov--Arnol'd--Moser (KAM) theorem, that shows, for a large enough class of dynamical systems, that regular motion can be recovered when the strength of the perturbation is taken formally to go at infinity. Invariant tori reappear after the system crossed regions of chaotic motion. In this way, we extend our previous work in this direction obtaining firmer mathematical results. Applications to the double pendulum and a re-analysis of the harmonic oscillator in a traveling wave are also provided to show the strength of this approach in practical examples.
\end{abstract}
\maketitle
\raggedbottom

\section{Introduction}
In nonlinear dynamics, it is essential to understand the behavior of the system as the perturbation increases for the onset of chaotic behavior. This is generally determined by a perturbation parameter that can determine the leading dynamics. Chirikov criterion of overlapping resonances \cite{Chirikov1979} is rather useful to fix a stochasticity threshold. This by-no-means implies that chaos can increase monotonically with the given increasing parameter far beyond the regime in which the original resonances were identified. With a very large parameter (strong coupling) different scales could emerge where dynamics turns out to be regular with a reconstitution of invariant tori.

The KAM theorem \cite{Kolmogorov1954,Arnold1963,Moser1962} establishes a rigorous foundation for understanding the effect of small perturbations on integrable Hamiltonian systems, demonstrating that a large measure of non-resonant invariant tori persists under suitable regularity and non-degeneracy assumptions. Specifically, the measure of these preserved tori approaches full measure as the perturbation parameter tends to 0. A central role in the proof is played by the control of small divisors arising in the perturbative construction. Extending these results to the strongly perturbed regime requires identifying which component of the system dominates the dynamics, thereby allowing an effective reformulation to which KAM theory can still be properly applied, given the proper regularity and non-degeneracy assumptions are fulfilled.

Dual perturbation theory approaches this problem by interchanging the roles of the leading and perturbing terms. An early formulation of this principle related inverse-parameter expansions to the quantum adiabatic approximation \cite{Frasca1998}. The classical dual Lindstedt construction of Ref.~\cite{Frasca2009} extended the connection of this technique with KAM theory. In this latter case, time and momentum rescaling reveal a strongly coupled perturbation problem around which a new expansion can be organized. Our aim in this article is to put on firmer mathematical ground the results presented in Ref.~\cite{Frasca2009}, adding an extended analysis of applications.

Current quantitative KAM estimations emphasize the domain issue. The measure excluded for nondegenerate systems from the primary tori goes like the square root of the perturbation size with constants depending on the integrable system and the observation domain \cite{BiascoChierchia,ChierchiaKoudjinan}. Thus, we should should for a proper rescaling, in our formal limit of large perturbation, that goes like some the reciprocal of the square root of the perturbation size. We point out that large perturbation can also be treated by more specialized techniques (e.g., for Duffing oscillators see \cite{YuanChenLi}).

Another relevant aspect is the cancellations among terms in the Lindstedt expansion for its convergence. This mechanism in configuration space was studied in \cite{CorsiGentileProcesi}. We will not address this point in this article.

From the examples we consider in this work, it will appear clear enough as a relation should exist between our approach and some well-known averaging and singular perturbation methods \cite{Sanders2007}. The interplay between the time-scales is a relevant clue for this connection but it should be said that e.g. secularities occur in the strong perturbation expansion as well and should be properly treated exactly in the same way as for the weak perturbation case. Similarly, the question of separatrix-crossing and nearing singularities should be addressed adequately. We know that separatrix passages can change adiabatic invariants \cite{Cary1986}. 

We combine an exact geometric estimate with an explicit local perturbative calculation in our study. In our applications, we address strong coupling up to $10^2$ in the simplest case of the double pendulum and $10^4$ for the oscillator in a traveling wave that, together with width measurements, and independent short-segment integrations check the predictions with respect to a strong perturbation.

The paper is organized as follows. In Sec. II--IV we establish the formalism, physical domains, and autonomous measure estimate, including its dependence on the kinetic energy and remainder. In Sec. V we treat the double pendulum case. In Sec. VI we develop the study of the traveling-wave model. Finally, Sec. VII yields the conclusions.

\section{Exact scaling and physical domains}

Let $n\geq2$ the dimensions of the configuration space and $\lambda>0$ the size of the perturbation, and consider
\begin{equation}
 H_\lambda(p,q)=\frac12p^{\mathsf T}M^{-1}p
                 +\lambda U(q)+W(q),
 \label{eq:H}
\end{equation}
where $M$ is a constant positive-definite mass matrix. Both $U(q)$ and $W(q)$ are potentials acting on the system. The configuration space may be a torus or a region of $\R^n$ but only the regular phase-space domain specified below is needed. Define
\begin{equation}
 p=\sqrt\lambda P,\qquad \tau=\sqrt\lambda t,
 \qquad S_\lambda(P,q)=(\sqrt\lambda P,q).
 \label{eq:scale}
\end{equation}
Here, $S_\lambda$ is a map restoring the physical variables. In this case, we have just a rescaling in momenta. Hamilton's equations in $(P,q,\tau)$ are exactly those generated by
\begin{equation}
 K_\lambda(P,q)=\lambda^{-1}H_\lambda(\sqrt\lambda P,q)
 =K_\infty(P,q)+\lambda^{-1}W(q),
 \label{eq:K}
\end{equation}
with $K_\infty=P^{\mathsf T}M^{-1}P/2+U(q)$.

The momentum map alone is not canonical: if
$\Omega=\sum_i dp_i\wedge dq_i$, then
$S_\lambda^*\Omega=\sqrt\lambda\sum_i dP_i\wedge dq_i$.
Its constant conformal factor is compensated by the time and Hamiltonian scalings. In particular, it maps Lagrangian invariant tori to Lagrangian invariant tori. For the respective flows,
\begin{equation}
 \Phi_{H_\lambda}^{t}\circ S_\lambda
 =S_\lambda\circ\Phi_{K_\lambda}^{\sqrt\lambda t}.
 \label{eq:flows}
\end{equation}
This identity grants that the flows are invariant under this rescaling even if the momentum rescaling is not canonical.

Suppose a canonical analytic chart $\Psi:(I,\theta)\mapsto(P,q)$ puts $K_\infty$ into the form $h(I)$ on a regular region. For a bounded action domain $D$ of positive volume and phases in the domain $\T^n$, set
\begin{equation}
 \mathcal A=\Psi(D\times\T^n),\qquad
 \mathcal A_\lambda=S_\lambda\mathcal A.
 \label{eq:domains}
\end{equation}
The physical domains expand in momentum, and
\begin{equation}
 \meas(\mathcal A_\lambda)=\lambda^{n/2}\meas(\mathcal A).
 \label{eq:jac}
\end{equation}
All measure statements below refer to Liouville phase volume on these domains rather than to a microcanonical measure on an energy surface.

A fixed scaled datum $(P_0,q_0)$ corresponds to
$p_0(\lambda)=\sqrt\lambda P_0$. A fixed physical datum instead corresponds to $P_0(\lambda)=p_0/\sqrt\lambda$ and need not remain in $\mathcal A$. Consequently, an ensemble theorem on $\mathcal A_\lambda$ does not decide the fate of each prescribed physical initial condition. There is no contradiction between almost-full relative measure on these expanding domains and exceptional trajectories outside them.

\section{A quantitative strong-coupling corollary}

We assume that $\lambda$ is just a signpost for ordering and it is not expected to appear otherwise. Let $D$ be a bounded action domain with $C^2$ boundary and compact closure in a larger open domain $D_+$. We assume the following hypotheses

\begin{enumerate}
\item The canonical chart $\Psi$ extends analytically to a fixed complex neighborhood of $\overline{D_+}\times\T^n$, with positive action radius $r$ and angular strip width $s$.
\item On this neighborhood $K_\infty\circ\Psi=h(I)$ and $f=W\circ\Psi$ are analytic and real on the real domain. Their relevant analytic norms are finite, and $\norm{f}_{r,s}\leq F$.
\item The frequency map $\omega(I)=\partial_Ih(I)$ is injective on $D_+$, and its derivative is uniformly invertible there: the Hessian and its inverse have finite bounds on the real domain. The complex extensions and boundary margins are fixed.
\end{enumerate}
We do not discuss the case of finitely many frequency charts, this is the reason why we assume injectivity. Constants can depend on fixed domains, their geometry, the indexes $n,r,s$, and analytic and twist bounds. 

The standard analytic KAM measure theorem in this setting gives a union of invariant Lagrangian tori whose complement in $D\times\T^n$ has measure bounded by $C\sqrt\delta$ when the analytic perturbation norm is at most $\delta$ and $\delta$ is sufficiently small \cite{BiascoChierchia,ChierchiaKoudjinan}. The surrounding domain $D_+$ accommodates the displaced tori while the bound counts points in the fixed observation domain $D\times\T^n$.

\begin{theorem}[Strong-coupling measure bound]\label{theo1}
Under the preceding assumptions there are constants $\delta_*>0$ and $C_D>0$ such that, for $F/\lambda<\delta_*$, a union $\mathcal T_\lambda$ of invariant Lagrangian tori of $H_\lambda$ satisfies
\begin{equation}
 \frac{\meas(\mathcal A_\lambda\setminus\mathcal T_\lambda)}
      {\meas(\mathcal A_\lambda)}
 \leq C_D\sqrt{\frac{F}{\lambda}}.
 \label{eq:measure}
\end{equation}
The tori are defined in a neighborhood of $\mathcal A_\lambda$ and carry quasiperiodic dynamics. Their physical frequency vectors are $\sqrt\lambda$ times their frequency vectors in scaled time.
\end{theorem}

\begin{proof}
In action--angle variables Eq.~\eqref{eq:K} reads
\begin{equation}
 K_\lambda\circ\Psi=h(I)+\lambda^{-1}f(I,\theta).
\end{equation}
Its perturbation norm is at most $\delta=F/\lambda$. The quoted analytic KAM theorem applies with constants independent of $\lambda$. For a fixed Diophantine exponent $\nu>n-1$, its smallness condition can be arranged as $\delta\leq c\gamma^2$, with
\begin{equation}
 |k\cdot\omega_*|\geq\gamma|k|_1^{-\nu},
 \qquad 0\ne k\in\mathbb Z^n.
 \label{eq:dio}
\end{equation}
Taking $\gamma=A\sqrt\delta$ with a sufficiently large fixed $A$ gives the square-root measure bound. The canonical chart preserves volume. The map $S_\lambda$ multiplies both the excluded volume and the observation volume by the same factor $\lambda^{n/2}$, so their ratio is unchanged. Equation~\eqref{eq:flows} transfers invariance and rescales frequencies. The case $F=0$ is the integrable case and is immediate.
\end{proof}

Values for $C_D$ or $\delta_*$ can in principle evaluated by explicit KAM esimates and certified bounds on the chart. Eq.~\eqref{eq:measure} represents a quantitative asymptotic statement.

At fixed $F$, the absolute excluded volume is bounded only by
\begin{equation}
 \meas(\mathcal A_\lambda\setminus\mathcal T_\lambda)
 \leq C_D\meas(\mathcal A)\sqrt F\,
             \lambda^{(n-1)/2}.
 \label{eq:absolute}
\end{equation}
The relative bound tends to zero, whereas this upper bound on absolute volume grows for $n\geq2$. The volume of the total domain increases while excluded fraction tends to shrink even more.

\section{Momentum dependence and alternative scaling}

This argument also generalizes in the following case
\begin{equation}
 K_\lambda\circ\Psi=h(I)+R_\lambda(I,\theta),\qquad
 \norm{R_\lambda}_{r,s}\leq B\lambda^{-\beta},
 \label{eq:remainder}
\end{equation}
with $\beta>0$ and uniform analytic and twist hypotheses. No analyticity in $1/\lambda$ is required.

\begin{corollary}
For the family \eqref{eq:remainder}, the relative excluded measure in the corresponding physical observation domains is $\OO(\lambda^{-\beta/2})$.
\end{corollary}
\begin{proof}
Apply the preceding KAM estimate with $\delta=B\lambda^{-\beta}$ and cancel the constant Jacobian of the scaling in the relative measure.
\end{proof}

For example, consider a subdominant momentum-dependent term,
\begin{equation}
 H_\lambda=\tfrac12p^{\mathsf T}M^{-1}p
                     +\lambda U(q)+W(p,q).
\end{equation}
The residual term is $\lambda^{-1}W(\sqrt\lambda P,q)$. If its pullback to the fixed complex action--angle domain obeys
\begin{equation}
 \norm{W(\sqrt\lambda P,q)\circ\Psi}_{r,s}
 \leq B\lambda^\sigma,\qquad \sigma<1,
\end{equation}
then $\beta=1-\sigma$. 

When we consider a polynomial momentum of degree $d$, with analytic coefficients uniformly bounded on the chart, $\sigma=d/2$ is a sufficient growth exponent. E.g., degree zero gives $\beta=1$ and degree one gives $\beta=1/2$. A quadratic term generally remains of order one and must be incorporated into the dominant Hamiltonian if integrability and twist survive. If there are terms of degree greater than two, they invalidate such a perturbative split. For example, $W=p_1^4$ produces $\lambda P_1^4$ after normalization that is not manageable perturbatively. Rather, a different scaling should be devised to entering the validity assumptions of Theorem~\ref{theo1}.

If the dominant potential itself depends on momentum, then
\begin{align}
 H_\lambda&=\tfrac12p^{\mathsf T}M^{-1}p
               +V_0(p,q)+\lambda V_1(p,q),\\
 K_\lambda&=\tfrac12P^{\mathsf T}M^{-1}P
        +V_1(\sqrt\lambda P,q)
        +\lambda^{-1}V_0(\sqrt\lambda P,q).
 \label{eq:general}
\end{align}
Integrability of an expression with unscaled arguments does not supply uniform control of Eq.~\eqref{eq:general}. 

The quadratic kinetic term motivates the square-root momentum scaling. Instead, let us consider the following case
\begin{equation}
 H_\lambda=T(p)+\lambda U(q)+W(q),\qquad
 T(cp)=c^rT(p),\quad c>0,
\end{equation}
where $r>1$ and all functions are analytic on the domains used. A proper scaling is
\begin{equation}
 p=\lambda^{1/r}P,\qquad
 \tau=\lambda^{1-1/r}t,
\end{equation}
and the normalized Hamiltonian becomes
$K_\lambda=T(P)+U(q)+\lambda^{-1}W(q)$.
If this dominant Hamiltonian satisfies the same action--angle and twist assumptions, the relative bound remains $\OO(\lambda^{-1/2})$. The physical volume factor is now $\lambda^{n/r}$ and the frequency factor is $\lambda^{1-1/r}$. Homogeneity alone does not imply integrability, and a noninteger power requires care about analyticity near $P=0$.

\section{Application to two coupled pendula}

Consider, in dimensionless units,
\begin{align}
 H_\lambda={}&\frac{p_1^2+p_2^2}{2}
 -\lambda(a_1\cos q_1+a_2\cos q_2)\nonumber\\
 &+\kappa\cos(q_1-q_2),\qquad a_i>0.
 \label{eq:pendula}
\end{align}
The dominant scaled Hamiltonian is a sum of two independent pendulum Hamiltonians,
\begin{equation}
 K_\infty=\sum_{i=1}^2\left(\frac{P_i^2}{2}-a_i\cos q_i\right),
\end{equation}
and the perturbation parameter is $\eta=\kappa/\lambda$.

For one pendulum $P^2/2-a\cos q$ on a librational orbit, write its energy as
\begin{equation}
\label{eq:e}
 e=-a+2ak^2,\qquad 0<k<1,
\end{equation}
where $k=\sin(q_{max}/2)$, being $q_{max}$ the libration maximum angle. Using the elliptic integrals $\mathrm K(k)$ and $\mathrm E(k)$, with $k$ the modulus, we can write action and frequency as
\begin{align}
 I(k)&=\frac{8\sqrt a}{\pi}
 [\mathrm E(k)-(1-k^2)\mathrm K(k)],\label{eq:action}\\
 \omega(k)&=\frac{\pi\sqrt a}{2\mathrm K(k)}.
 \label{eq:freq}
\end{align}
The identities derived in the appendix give
\begin{equation}
 \frac{dI}{dk}=\frac{8\sqrt a}{\pi}k\mathrm K(k)>0,
 \qquad
 \frac{d\omega}{dI}
 =-\frac{\pi^2\mathrm K'(k)}{16k\mathrm K(k)^3}<0.
 \label{eq:twist}
\end{equation}
We can choose fixed closed modulus intervals
\begin{equation}
 0<k_{i,-}\leq k_i\leq k_{i,+}<1
 \label{eq:moduli}
\end{equation}
and a bounded smooth action domain $D$ lying strictly inside the corresponding product of action intervals. The pendulum action--angle charts are analytic on a common complex neighborhood after the widths have been reduced if necessary. The Hessian of
$h(I_1,I_2)=e_1(I_1)+e_2(I_2)$, with $e_1(I_1)$ and $e_2(I_2)$ given by eq.(\ref{eq:e}), is diagonal with strictly negative diagonal entries as $I_i$ depends non-trivially from $k_i$, bounded away from zero on these compact intervals. Its frequency map is injective there. Thus all assumptions of Theorem~\ref{theo1} hold. The restriction $k_{i,-}>0$ avoids the coordinate singularity of action--angle variables at the elliptic equilibrium; $k_{i,+}<1$ avoids the separatrix.
\begin{proposition}
For every fixed domain specified above, there are constants $C$ and $\eta_*>0$ such that, if $|\kappa|/\lambda<\eta_*$, the physical system \eqref{eq:pendula} has a family of invariant Lagrangian tori satisfying
\begin{equation}
 \frac{\meas(\mathcal A_\lambda\setminus\mathcal T_\lambda)}
 {\meas(\mathcal A_\lambda)}
 \leq C\sqrt{\frac{|\kappa|}{\lambda}}.
 \label{eq:pendbound}
\end{equation}
\end{proposition}
\begin{proof}
In the product pendulum chart, the analytic norm of
$\cos(q_1-q_2)$ is bounded by a constant $B_D$ on a fixed complex neighborhood. Its complex norm need not be at most one, even though its real value is. Apply Theorem~1 with $F=|\kappa|B_D$ and absorb $B_D$ into the constants.
\end{proof}

This is true also when $a_1=a_2$. The diagonal resonance $\omega_1=\omega_2$ does not make the Hessian singular. Such a resonance is excluded from the primary Diophantine set. For the question of the separatrix recovery, we do not address this problem here.

\begin{figure*}[t]
\centering
\includegraphics[width=.88\textwidth]{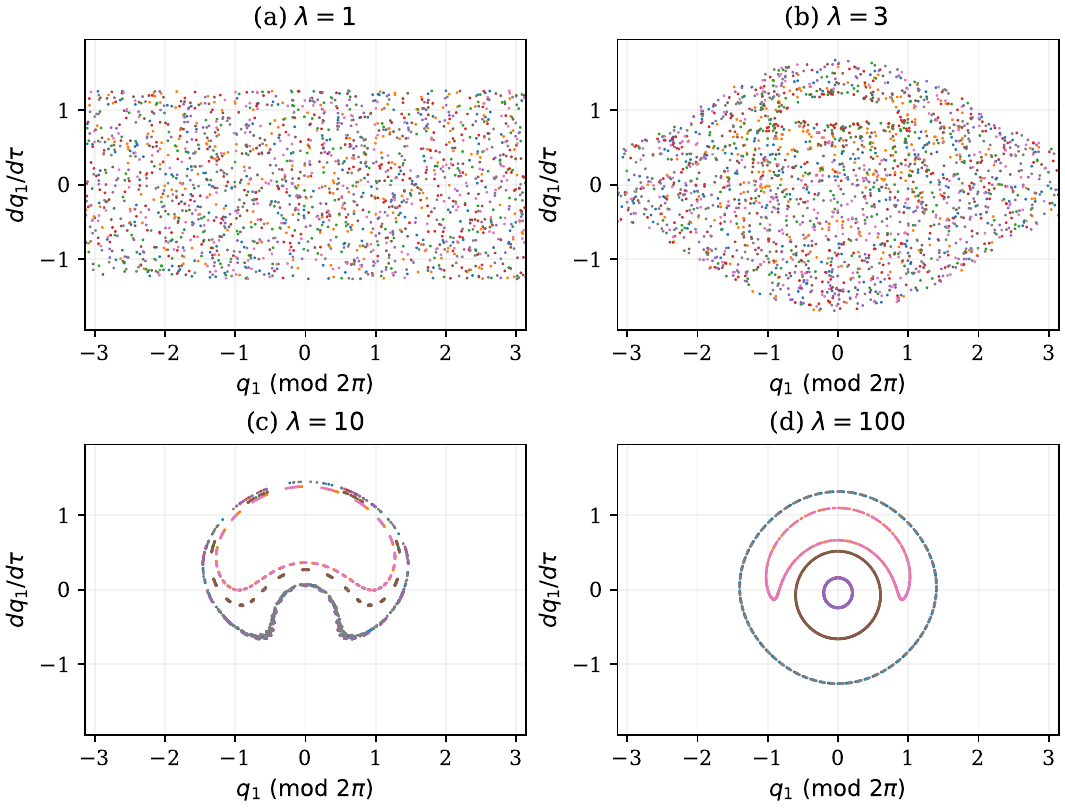}
\caption{Poincare sections of two coupled pendula with $a_1=1$, $a_2=1.3$, $\kappa=1$, and scaled mechanical energy $\mathcal E=-0.5$. The section is $q_2=0\pmod{2\pi}$ with $q_2'>0$; $q_1$ is displayed in $[-\pi,\pi)$. Eight initial conditions per panel have $q_2(0)=q_1'(0)=0$, $q_1(0)=\pm0.2,\pm0.6,\pm1.0,\pm1.4$, and positive $q_2'(0)$.}
\label{fig:pend}
\end{figure*}

\section{The traveling-wave model: a singular forcing limit}

We now consider the same time-dependent model analyzed in Ref.~\cite{Frasca2009}
\begin{equation}
H_\lambda(p,q,t)=\frac{p^2}{2}+\frac{\omega_0^2q^2}{2}
                  -\lambda\alpha\cos(q-\omega t),
 \label{eq:wave}
\end{equation}
where $\alpha>0$, $\omega_0>0$, and $\omega\ne0$ are fixed. Define $\eps=1/\lambda$ and $\nu=\omega\sqrt\eps$. The exact scaled system is
\begin{equation}
 K_\eps(P,q,\tau)=\frac{P^2}{2}-\alpha\cos(q-\nu\tau)
                       +\frac{\eps\omega_0^2q^2}{2}.
 \label{eq:wavescaled}
\end{equation}
Its forcing period in $\tau$ is $2\pi/|\nu|$. The behavior of such a system can be understood without Hamiltonian methods. Without rescaling time, Newton's equation is
\begin{equation}
 \ddot q+\omega_0^2q+\lambda\alpha\sin(q-\omega t)=0.
 \label{eq:newtonwave}
\end{equation}
For $\lambda=0$ it is the harmonic oscillator, with general solution $q=A\cos(\omega_0t)+B\sin(\omega_0t)$. If instead the harmonic force is omitted, set $x=q-\omega t$. Then
\begin{equation}
 \ddot x+\lambda\alpha\sin x=0,\qquad
 \tfrac12\dot x^2-\lambda\alpha\cos x=\text{constant}.
 \label{eq:purewave}
\end{equation}
As expected, the motion in a single traveling plane-wave potential is exactly integrable. The first integral follows simply by multiplying the equation by $\dot x$. This is the positive content of the interpolation argument: the two dominant descriptions are integrable, although they describe different orbits and different natural scales.

For the complete equation, introduce $\tau=\sqrt\lambda t$ and $Q=q-\omega t=q-\nu\tau$, where primes denote derivatives with respect to $\tau$. Direct differentiation gives
\begin{equation}
 Q''+\alpha\sin Q=-\eps\omega_0^2 Q
                         -\eps\omega_0^2\nu\tau.
 \label{eq:movingforce}
\end{equation}
For fixed scaled initial data and on every fixed compact $\tau$ interval, this converges to the pendulum equation as $\eps\to0$. Continuous dependence of solutions follows from the locally uniform convergence of the force and its Lipschitz bound for finite intervals. For fixed physical initial velocity $v_0$, the scaled initial velocity is $Q'(0)=(v_0-\omega)/\sqrt\lambda$, which also has a definite limit. 

At large $\lambda$, harmonic confinement can be neglected on bounded spatial domains and fixed intervals of rescaled time. If $q$ grows in proportion to $\lambda$, the neglected force can become of order one. In the wave frame, the time-dependent correction also becomes of order one on the physical time scale $t\sim\lambda$. This marks the limit of the force estimate and, by itself, establishes neither an escape time nor a threshold for chaos.

At any finite $\lambda$, the particle cannot follow the same moving potential well indefinitely. If it did, the harmonic restoring force would grow without bound, while the wave force would remain bounded. Regular motion in the laboratory frame remains possible.

At $\lambda=0$, the harmonic oscillator has the same frequency at every amplitude and therefore lacks the twist required by the KAM theorem used here. The two integrable limits suggest a return to regular motion at strong coupling, but whether a regular–chaotic–regular sequence occurs depends on the initial conditions and requires separate verification.

\begin{figure*}[t]
\centering
\includegraphics[width=.88\textwidth]{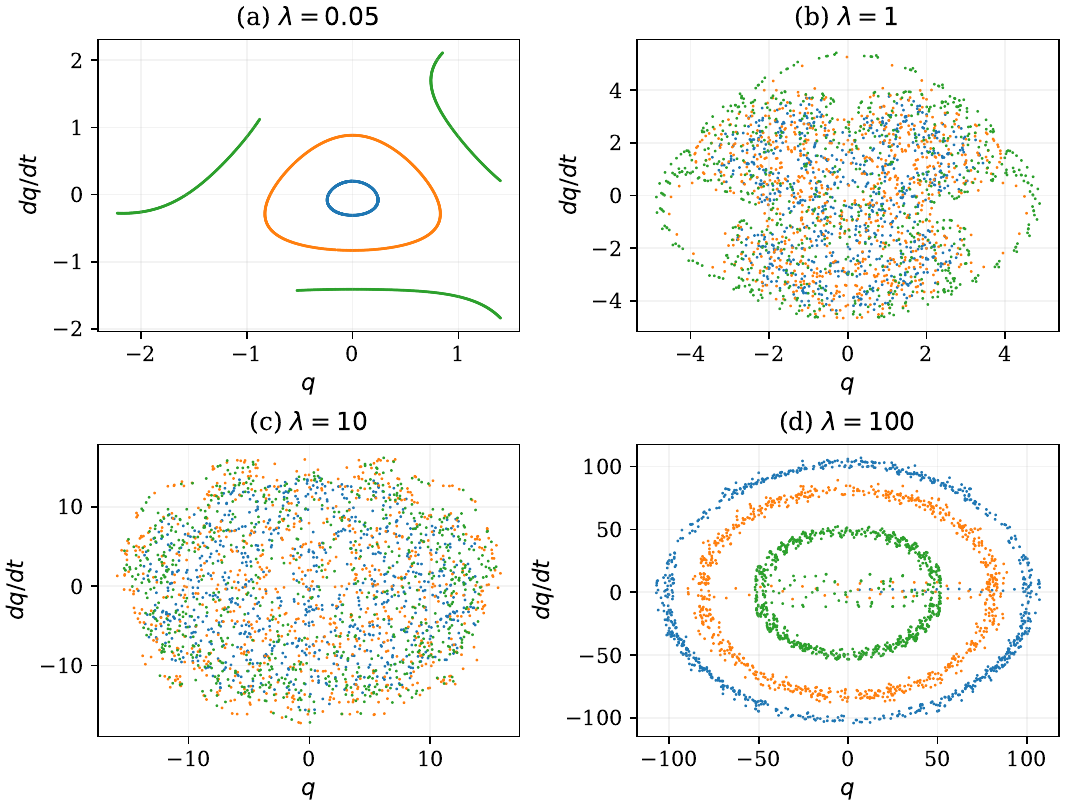}
\caption{Stroboscopic sections of the full Newton equation \eqref{eq:newtonwave}, with $\omega_0=\alpha=1$ and $\omega=1.5$. Each panel uses the same three physical initial conditions: $\dot q(0)=0.2$, and $q(0)=0,0.7,1.4$ (blue, orange, and green). For each orbit 800 samples at $t_n=2\pi n/\omega$, $n=1,\ldots,800$, are shown, without discarding an initial interval.}
\label{fig:wave}
\end{figure*}

In Fig.~\ref{fig:wave}, we show how the dynamics changes at increasin $\lambda$.
The nearly annular geometry in Fig.~\ref{fig:wave}(d) may suggest that some diffuse region still persists even if it become even more smaller at values of $\lambda$ even more greater. Indeed, the bands become thinner. This can be shown as follows. Write $\Lambda=\lambda\alpha$ and define the harmonic energy and its elliptical radius by
\begin{equation}
 E_h=\tfrac12(\dot q^2+\omega_0^2q^2),\qquad
 r=\sqrt{q^2+(\dot q/\omega_0)^2}.
 \label{eq:ellipradius}
\end{equation}
Level sets of $r$ are exactly harmonic-oscillator ellipses, with semiaxes $r$ and $\omega_0r$. Multiplying Newton's equation by $\dot q$, and integrating the wave term by parts, gives the exact identity
\begin{align}
 E_h(t)-E_h(s)={}&\Lambda[\cos(q(t)-\omega t)
                       -\cos(q(s)-\omega s)]\nonumber\\
 &-\Lambda\omega\int_s^t\sin(q(u)-\omega u)\,du.
 \label{eq:harmwork}
\end{align}
Consequently, on any time interval $J$ of duration $W$,
\begin{equation}
 |r^2(t)-r^2(s)|\leq
 B_\lambda(W):=\frac{2\lambda\alpha}{\omega_0^2}
                         (2+|\omega|W),\quad s,t\in J.
 \label{eq:radiusbound}
\end{equation}
The following proposition can be proven.
\begin{proposition}[Finite-window harmonic geometry]
Suppose an orbit reaches, at a possibly $\lambda$-dependent time $t_\lambda$, a radius $R_\lambda=r(t_\lambda)$ with $R_\lambda/\lambda\to c>0$. On $J_\lambda=[t_\lambda,t_\lambda+W]$, let $r_-$ and $r_+$ be the minimum and maximum radius. Then
\begin{equation}
 \Delta r:=r_+-r_-\leq
 \frac{B_\lambda(W)}{r_++r_-}.
 \label{eq:fullwidth}
\end{equation}
For fixed $W$, $r_\pm=R_\lambda+\OO_W(1)$, so
\begin{equation}
 \Delta r=\OO_W(1),\qquad
 \frac{\Delta r}{R_\lambda}=\OO_W(\lambda^{-1}).
 \label{eq:relativewidth}
\end{equation}
More generally, if $W=W_\lambda=o(\lambda)$, then
$\sup_{J_\lambda}|r/R_\lambda-1|\to0$.
\end{proposition}
\begin{proof}
Apply Eq.~\eqref{eq:radiusbound} to times attaining the radial extrema and factor $r_+^2-r_-^2$. Applying the same inequality with $s=t_\lambda$ gives $|r^2-R_\lambda^2|\leq B_\lambda(W)$. Since $R_\lambda^2\sim c^2\lambda^2$, division by $R_\lambda^2$ and taking square roots prove both claims. In particular, for fixed $W$ the bound in Eq.~\eqref{eq:fullwidth} is asymptotically no larger than $\alpha(2+|\omega|W)/(\omega_0^2c)$, up to $o(1)$.
\end{proof}

The conclusion is geometric: the trajectory lies in a increasingly thinner elliptical annulus as $\lambda$ increases. Large radii are observed numerically for the seeds considered here. Indeed, the forced large-excursion branch has $q,\dot q$ of order $\lambda$.

\begin{figure*}[t]
\centering
\includegraphics[width=.88\textwidth]{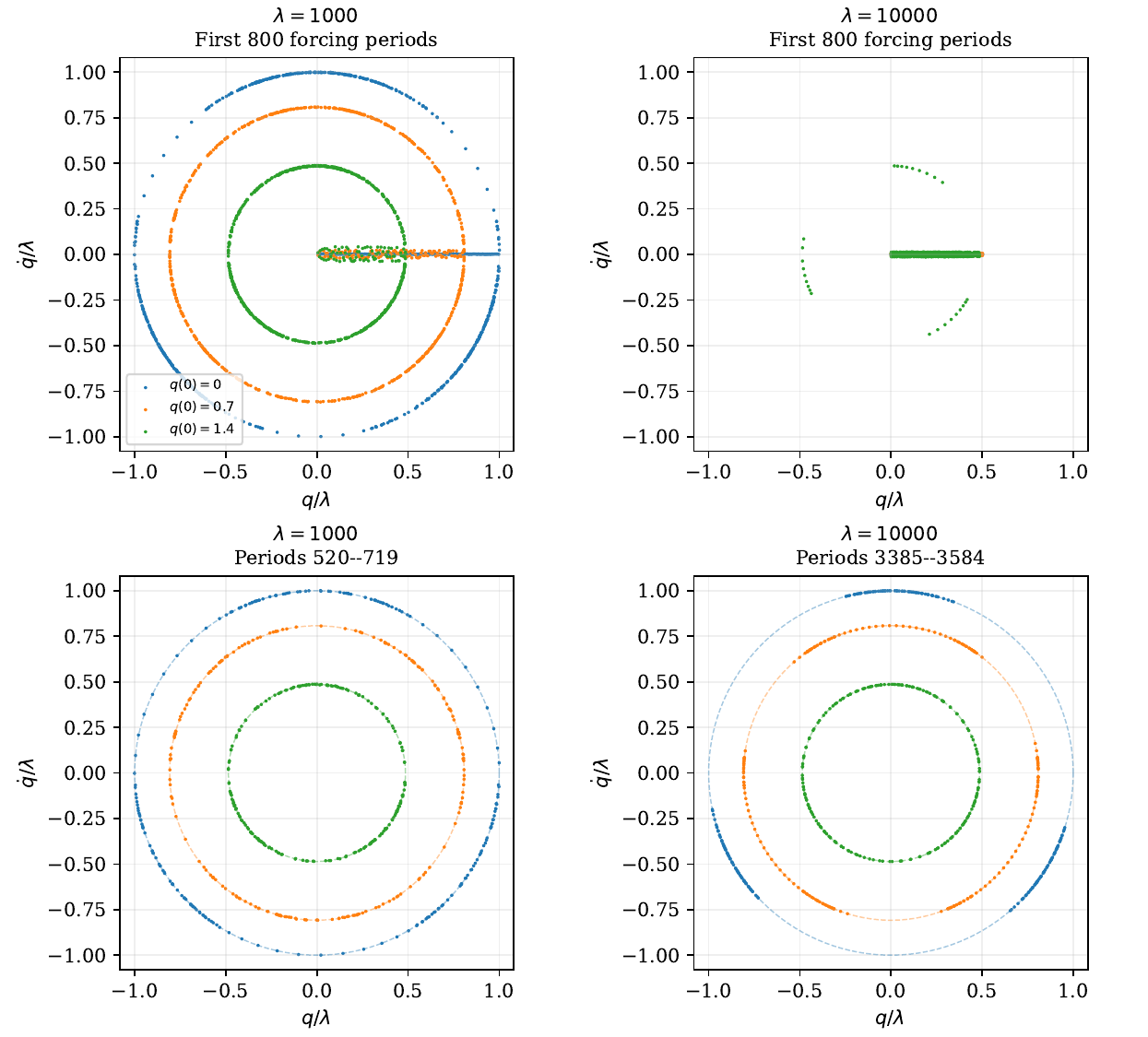}
\caption{Newtonian traveling-wave sections at $\lambda=1000$ and $10000$, in normalized coordinates. The first 800 periods and the late second window are displayed separately, as labeled. Colors identify the same seeds as Fig.~\ref{fig:wave}. Dashed harmonic circles are references. The late windows expose the thin large-excursion bands; the early window at $10000$ also contains the approach to this regime.}
\label{fig:large}
\end{figure*}

In order to see this behavior, we extend the Newtonian integrations to $\lambda=1000$ and $10000$, retaining the same physical seeds and the same two late windows. Figure~\ref{fig:large} separates the first 800 samples from the second late window. At $\lambda=10000$, the first 800 periods still include substantial drift; comparing that interval alone would mix the approach to the large-excursion branch with its subsequent geometry. The inner-band relative width is seen to decrease from $0.915\%$ to $0.0850\%$ between these couplings. 

In the wave frame, at fixed scaled time $\tau=\sqrt{\lambda}\,t$, the strong-coupling limit gives a pendulum as long as $q$ stays bounded. The harmonic force in the scaled equation is proportional to $q/\lambda$, so it vanishes in that limit. The harmonic oscillator describes a different regime. Once the orbit reaches a radius proportional to $\lambda$, confinement can remain important. On segments where the particle crosses many wave crests rapidly, successive pushes from the wave largely cancel, and the motion in the laboratory frame closely follows a harmonic orbit. So, let $q_*$ be the harmonic solution initialized at the \emph{exact} position and velocity at $t_0$. Assume $R_\lambda=r(t_0)\sim c\lambda$ and that on $J=[t_0,t_0+W]$ its velocity obeys $|\dot q_*-\omega|\ge d_0\lambda$, with $W,d_0,c>0$ fixed. Then for sufficiently large $\lambda$,
\begin{equation}
 \sup_J\left(|q-q_*|+|\dot q-\dot q_*|/\omega_0\right)
 =\mathcal O_W(1).
 \label{eq:passbound}
\end{equation}
Constants may depend on $d_0,c$ and the fixed physical parameters. Dividing by $R_\lambda$ gives an $\mathcal O_W(\lambda^{-1})$ bound in normalized phase space. The angular separation of the two radius vectors is consequently $\mathcal O_W(\lambda^{-1})$ throughout the segment.

\begin{proof}
Equation~\eqref{eq:radiusbound} implies $|q|=\mathcal O_W(\lambda)$, hence Newton's equation gives $|\ddot q|=\mathcal O_W(\lambda)$. Put $\psi=q-\omega t$ and $g=\dot q-\omega$. Initially $|g|\ge d_0\lambda$. Up to a possible first time when $|g|=d_0\lambda/2$, variation of constants expresses each component of the difference in Eq.~\eqref{eq:passbound} as $-\Lambda\int_{t_0}^{t}K(t-u)\sin\psi(u)\,du$, with kernels $\sin(\omega_0 s)/\omega_0$ and $\cos(\omega_0s)$ for position and velocity. Both kernels and their derivatives are bounded. Substitute $\sin\psi=-(\cos\psi)'/g$ and integrate by parts. Each endpoint term is $\mathcal O(\lambda^{-1})$ before multiplication by $\Lambda$. The integral has the same order because $1/|g|=\mathcal O(\lambda^{-1})$, $|\dot g|/g^2=\mathcal O_W(\lambda^{-1})$, and $W$ is fixed. Thus the velocity difference remains bounded independently of $\lambda$. At the proposed first time it would be at least $d_0\lambda/2$ by the assumed gap for $\dot q_*$. This contradiction extends the estimate to all of $J$. In coordinates $(q,\dot q/\omega_0)$ the harmonic radius equals $R_\lambda$; two vectors separated by bounded distance at this radius subtend an angle of order $1/R_\lambda$.
\end{proof}

The bound limits the phase error over the segment. The perturbative correction describes the leading oscillation about the harmonic motion.

The harmonic limit on this branch admits a constructive perturbative description. Put $s=t-t_0$ and choose a harmonic carrier
\begin{align}
 q_h(s)&=Q\cos(\omega_0s)+\frac{V}{\omega_0}\sin(\omega_0s),\\
 v_h&=\dot q_h,\quad a_h=-\omega_0^2q_h,\\
 \psi_h&=q_h-\omega(t_0+s),\quad d=v_h-\omega.
\end{align}
On a fixed passing segment assume $R_h=\sqrt{Q^2+(V/\omega_0)^2}\asymp\lambda$ and $|d|\geq d_0\lambda$, with $d_0>0$ independent of coupling. Direct differentiation in Newton's equation gives the first oscillatory displacement
\begin{align}
 u_1&=\frac{\Lambda}{d^2}\sin\psi_h,\label{eq:uone}\\
 \dot u_1&=\frac{\Lambda}{d}\cos\psi_h
          -\frac{2\Lambda a_h}{d^3}\sin\psi_h.
\end{align}
Thus $u_1=\mathcal O(\lambda^{-1})$, whereas $\dot u_1=\mathcal O(1)$. The large forcing is balanced by the rapid second derivative of a small displacement. This explains how an order-one velocity corrugation can coexist with convergence of the normalized ellipse.

Substitution of $q_h+u_1$ leaves the leading oscillatory residual
\begin{equation}
 -\frac{3\Lambda a_h}{d^2}\cos\psi_h
 +\frac{\Lambda^2}{2d^2}\sin(2\psi_h)
 +\mathcal O(\lambda^{-1}).
\end{equation}
It is canceled by a further correction
\begin{equation}
 u_2=-\frac{3\Lambda a_h}{d^4}\cos\psi_h
       +\frac{\Lambda^2}{8d^4}\sin(2\psi_h),
 \label{eq:utwo}
\end{equation}
which has displacement order $\lambda^{-2}$ and velocity order $\lambda^{-1}$. The residual of $q_h+u_1+u_2$ is formally $\mathcal O(\lambda^{-1})$ under the stated segment assumptions. These are oscillatory outer corrections, not a claim of a complete uniform second-order solution: remaining slow terms can accumulate a displacement of order $\lambda^{-1}$ on a fixed segment. Consequently adding $u_2$ need not improve the order of the position error. The fast phase depends on $\lambda$, so this construction is not an ordinary power series with parameter-independent coefficient functions.

Write $q_h=R_h\cos\theta_h$ and $v_h=-\omega_0R_h\sin\theta_h$. Linearization of radius and angle gives
\begin{align}
 \delta r_1&=\cos\theta_h\,u_1
                -\frac{\sin\theta_h}{\omega_0}\dot u_1,\\
 \delta\theta_1&=-\frac{1}{R_h}\left(
       \sin\theta_h\,u_1+\frac{\cos\theta_h}{\omega_0}\dot u_1\right).
\end{align}
In particular, away from velocity matching,
\begin{align}
 \delta r_1&=\frac{\Lambda}{\omega_0^2R_h}\cos\psi_h
                 +\mathcal O(\lambda^{-1}),\label{eq:radcorr}\\
 \delta\theta_1&=-\frac{\Lambda\cos\theta_h}{\omega_0R_hd}
                 \cos\psi_h+\mathcal O(\lambda^{-2}).
\end{align}
Equation~\eqref{eq:radcorr} recovers the band-envelope estimate from the work identity and supplies its oscillatory mechanism. It predicts absolute corrugation of order unity and relative corrugation of order $\lambda^{-1}$. The angle correction is also of order $\lambda^{-1}$, although its instantaneous derivative need not be small.

\section{Conclusions}

The exact scaling permits a formal organization but does not, by itself, prove convergence of the resulting power series. Convergence mechanisms based on cancellations in Lindstedt series are analyzed in Ref.~\cite{CorsiGentileProcesi}; applying such results requires verifying their analytic and arithmetic hypotheses for the chosen reference dynamics. Our measure theorem uses KAM persistence directly and does not require an assertion that every formal Lindstedt series converges. Nor does it assign a torus to arbitrary prescribed initial data. In the traveling-wave problem a second scale $\nu=\omega\sqrt\eps$ enters, so a pure integer-power expansion with parameter-independent coefficients is not automatically available.

Strong-coupling torus recovery is rigorously supported when an exact scaling yields a uniformly analytic, nondegenerate integrable Hamiltonian plus a small remainder on a specified domain. For quadratic mechanical systems with a fixed configuration-dependent remainder, the relative measure not covered by the constructed tori is $\OO(\lambda^{-1/2})$ on physical domains whose momenta scale as $\sqrt\lambda$. Momentum dependence can change this rate or invalidate the chosen limit while homogeneous kinetic energies change the momentum and time scales without changing the relative square-root estimate when the normalized remainder is still $\OO(\lambda^{-1})$.

Two coupled pendula provide a direct application with explicitly verified twist on compact librational domains, and the Newtonian sections illustrate the restoration of organized motion at large coupling. Extending those domains toward separatrices requires additional estimates. For the traveling-wave model, Newton's equation confirms the interpolation between harmonic motion and an integrable pure-wave description. Its stroboscopic sections show a corresponding reorganization at strong coupling. 

The harmonic interpretation has a complementary quantitative basis: for large-excursion orbits with radius proportional to $\lambda$, a direct work bound yields relative radial thickness $\OO(\lambda^{-1})$ on fixed observation windows. This geometric conclusion is consistent with an absolute thickness of order unity, as observed on the inner branch. On fixed passing segments, integration by parts also proves an $\mathcal O_W(\lambda^{-1})$ phase bound relative to the harmonic solution with the same initial state. The first oscillatory correction resolves radial structure within the annulus. The tests up to $\lambda=10^4$ support the validity of this strong-coupling asymptotic mechanism. Infinite-time invariance, access to this branch, and matching through velocity-matching regions remain to be established.

\section*{Data availability}
No external datasets enter the numerical calculation. The Python scripts and numerical outputs needed to reproduce the figures are supplied on reasonable request to the author.

\section*{Use of AI tools}
ChatGPT Codex (GPT-6.2 Astra) has been used for literature review, calculations, some language editing and presentation. All derivations, numerical calculations, interpretations, and final text were checked and approved by the author.

\appendix*
\section{Pendulum action and twist identities}\label{app:twist}
For librational energy $e=-a+2ak^2$, the turning point satisfies $\sin(q_{\max}/2)=k$. The action is
\begin{equation}
 I=\frac{1}{2\pi}\oint P\,dq
   =\frac{2}{\pi}\int_0^{q_{\max}}\sqrt{2(e+a\cos q)}\,dq.
\end{equation}
With $\sin(q/2)=k\sin u$, the integral reduces to Eq.~\eqref{eq:action}, where
\begin{align}
 \mathrm K(k)&=\int_0^{\pi/2}\frac{du}{\sqrt{1-k^2\sin^2u}},\\
 \mathrm E(k)&=\int_0^{\pi/2}\sqrt{1-k^2\sin^2u}\,du.
\end{align}
Differentiation gives
\begin{equation}
 \frac{d}{dk}[\mathrm E-(1-k^2)\mathrm K]=k\mathrm K.
\end{equation}
Hence $dI/dk=8\sqrt a\,k\mathrm K/\pi$. Since $de/dk=4ak$, $de/dI=\pi\sqrt a/(2\mathrm K)$, and another differentiation gives Eq.~\eqref{eq:twist}. The integral defining $\mathrm K$ has positive derivative for $0<k<1$, proving the strict sign. As a consistency check, $\mathrm K(k)=\frac\pi2(1+k^2/4+\OO(k^4))$ implies
\begin{equation}
 e(I)=-a+\sqrt a\,I-\frac{I^2}{16}+\OO(I^3),
 \qquad h''(0)=-\frac18.
\end{equation}
The finite limiting twist does not remove the angular coordinate singularity at the center, which is why the stated domains exclude $I=0$.

\end{document}